\documentclass[pra,eqsecnum,twocolumn,floatfix,showpacs]{revtex4}

\usepackage{graphicx}
\usepackage{upgreek}
\usepackage{amsmath}
\usepackage{amssymb}
\usepackage{bm}

\newcommand{\vect}[1]{\mathbf{#1}}
\newcommand{\dyad}[1]{\mbox{\textbf{\textsf{#1}}}}

\newcommand{\be}{\begin{equation}}
\newcommand{\ee}{\end{equation}}
\newcommand{\tr}{\mathrm{Tr}}

\newcommand{\jsum}{\sum_{j=0}^\infty \!{}^{{}^\prime}}
\newcommand{\msum}{\sum_{m=0}^\infty \!{}^{{}^\prime}}

\newcommand{\dyG}{\dyad{G}}
\renewcommand{\th}{\tilde{h}}
\newcommand{\tj}{\tilde{\text{{\it \j}}}}
\newcommand{\tk}{\tilde{k}}
\newcommand{\ti}{\tilde{\text{{\it \i}}}}
\newcommand{\tlr}{\tilde{r}}
\newcommand{\ep}{\varepsilon}
\newcommand{\im}{\mathrm{Im}}
\newcommand{\re}{\mathrm{Re}}
\newcommand{\order}{\mathcal{O}}

\newcommand{\half}{{\textstyle\frac1{2}}}

\newcommand{\fourth}{{\textstyle\frac1{4}}}

\newcommand{\rMM}{r_\text{MM}}
\newcommand{\rNN}{r_\text{NN}}
\newcommand{\rMN}{r_\text{MN}}
\newcommand{\rNM}{r_\text{NM}}
\newcommand{\Meo}{\vect{M}_{{}^e_o m \eta}}
\newcommand{\Moe}{\vect{M}_{{}^o_e m \eta}}
\newcommand{\Neo}{\vect{N}_{{}^e_o m \eta}}
\newcommand{\Noe}{\vect{N}_{{}^o_e m \eta}}
\newcommand{\sincos}{\!\begin{array}{cc}\sin \\ \cos \end{array}\!}
\newcommand{\cossin}{\!\begin{array}{cc}\cos \\ \sin \end{array}\!}
\newcommand{\hr}{\hat{{\boldsymbol\varrho}}}
\newcommand{\hz}{\hat{\vect{z}}}
\newcommand{\hph}{\hat{\boldsymbol\varphi}}

\newcommand{\kB}{k_\mathrm{B}}
\newcommand{\op}{\omega_\mathrm{p}}
\newcommand{\mi}{i}

\newcommand{\Unr}{U^\mathrm{nr}}
\newcommand{\Ur}{U^\mathrm{res}}

\newcommand{\hankel}{{H^{(1)}_m}}

\newcommand{\comment}[1]{}

\begin{document}

\title{Casimir-Polder potential and transition rate in resonating
cylindrical cavities}

\begin{abstract}
We consider the Casimir-Polder potential of particles placed inside a 
metallic cylindrical cavity at finite temperatures, taking account of 
thermal non-equilibrium effects. In particular, we study how the
resonant (thermal non-equilibrium) potential and transition rates can
be enhanced by fine-tuning the radius of the cavity to match the
transition wavelength of the dominant transition of the particle.
Numerical calculations show that the CP potential acting atoms
prepared in low-lying Rydberg states can be enhanced beyond $30$kHz,
which is within the range of observability of modern experiments.
Because the magnitude of the resonance peaks depend sensitively on the
low frequency dissipation of the cavity metal, experiments in this
set-up could be a critical test of the disputed thermal
correction to the Casimir force between metal plates.
\end{abstract}

\date{\today}

\author{Simen {\AA}. Ellingsen}
\affiliation{Department of Energy and Process Engineering, Norwegian
University of Science and Technology, N-7491 Trondheim, Norway}
\author{Stefan Yoshi Buhmann}
\author{Stefan Scheel}
\affiliation{Quantum Optics and Laser Science, Blackett Laboratory,
Imperial College London, Prince Consort Road,
London SW7 2AZ, United Kingdom}

\pacs{
34.35.+a,  
12.20.--m, 
42.50.Ct,  
42.50.Nn   
}

\maketitle

\section{Introduction}

Casimir--Polder (CP) forces \cite{casimir48b} belong to the group of
of dispersion forces which arise due to the fluctuations of the
quantized electromagnetic field. They occur between polarizable atoms
or molecules and metallic or dielectric macroscopic bodies and can be
intuitively thought of as the dipole--dipole force caused by
spontaneous and mutually correlated polarization of the atom or
molecule and the matter comprising the body. Under the assumption of
thermal equilibrium, CP forces have been commonly investigated in the
linear-response formalism \cite{lifshitz55,mclachlan63,henkel02}. 

Recent theoretical predictions \cite{antezza05} as well as
experimental realizations \cite{obrecht07} for CP forces in thermal
non-equilibrium situations have pointed towards interesting effects
which arise when an atom at equilibrium with its local environment
interacts with a body held at a different temperature. In particular,
depending on the temperatures of the macroscopic body and the
environment, the force can change its character from being attractive
to repulsive and vice versa.

Non-equilibrium situations between atom and local environment can be
investigated by means of normal-mode quantum electrodynamics (QED)
\cite{nakajima97,wu00} or macroscopic QED in absorbing and dispersing
media \cite{buhmann07,scheel08}. In this case, thermal excitation and
de-excitation processes lead to resonant contributions to the force
\cite{buhmann08} (cf.~similar findings reported in
Ref.~\cite{gorza06}). At particle-body separations that are 
larger than the wavelengths associated with the dominating atomic
transitions (retarded regime), the interaction potential becomes 
spatially oscillating \cite{ellingsen09a}. Similar behaviour has been
observed for the transition rate of molecules in the past
\cite{drexhage74, chance78}.

The spatially oscillating non-equilibrium forces on ground-state
atoms or molecules are proportional to the thermal photon number.
In order to observe them, it is necessary to make use of 
atomic systems whose internal eigenstates exhibit energy separations
of order $\kB T$ or less ($T$: temperature, $\kB$: Boltzmann
constant). We have investigated polar molecules with their low-energy
rotational and vibrational transitions as possible candidates
\cite{ellingsen09a}. However, the large photon numbers obtained using
molecules with small excitation energies come at the cost: Due to the
large wavelengths associated with such molecular transitions, the
retarded regime where oscillations might be observed sets in at very
large distances, typically of the the order of tens to hundreds of
microns. As the CP potential decays away rapidly from the body
surface, it is very small at such distances. 

\begin{figure}[tb]
 \includegraphics[width=2.5in]{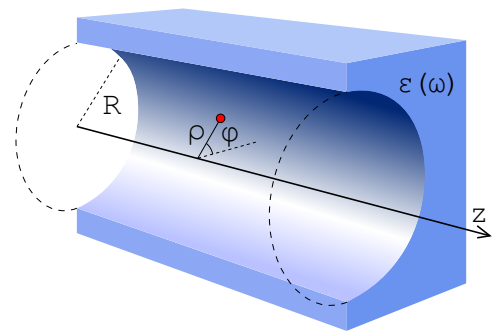}
 \caption{Cross-section of the geometry considered: a particle in a
 vacuum-filled circularly cylindrical cavity.}
 \label{fig:geom}
\end{figure}

As already discussed in conjunction with excited atoms in a zero
temperature environment, resonant forces can be enhanced in a planar
cavity whose width is fine-tuned to match the wavelength of the
transition
\cite{barton70,barton79,barton87,jhe91,jhe91b,hinds91,hinds94}.
Note that resonating cavities have been employed
experimentally for enhancement and inhibition of spontaneous emission
rates for excited systems for a long time (cf.\ e.g.\
\cite{kleppner81,goy83,unlu95,walther06}). We have previously
investigated the potential of such a setup to enhance the predicted
spatial oscillations of the thermal force on ground-state molecules
\cite{ellingsen09b}. Unfortunately, the cavity enhancement factor
turns out to scale logarithmically with the cavity $Q$ factor,
which strongly limits the possibilities of the scheme. As we have
shown, it is unlikely to achieve more than an order of magnitude's
enhancement of the force amplitudes, which is still insufficient for
detection using polar molecules.

In the present article, we present a scheme by which oscillations of
the resonant thermal CP force can be brought into the measurable
regime. This is achieved by replacing the planar cavity with a
cylindrical cavity \cite{ellingsen10} and employing Rydberg atoms
rather than polar molecules \cite{crosse10}. The geometry is shown in
Fig.~\ref{fig:geom}. The limited cavity enhancement in a planar setup
is due to the insufficient, purely one-dimensional confinement of the
electromagnetic modes. A cylindrical cavity is therefore an obvious
candidate for improvement: It confines the modes in two out of three
spatial dimensions, so that a stronger resonant enhancement may be
expected. At the same time, it is a practical geometry for
experimental and guiding purposes, allowing particles to travel freely
along the axial direction (we will only consider particles at rest
herein, leaving moving atoms \cite{scheel09} in this geometry for
future study). As a caveat, one has to bear in mind that a
cylindrical cavity will simultaneously enhance both the CP potential
and the relaxation rates; thus reducing the time scales on which
non-equilibrium effects can be observed. A similar geometry, i.e., an
(anisotropic) particle above a spherical hole in a thin metal plate,
has recently been suggested as a candidate for observing repulsive
Casimir forces \cite{levin10}.

Rydberg atoms are ideal candidates for observing resonant thermal CP
potentials. Their valence electrons are excited to relatively stable
states with very large principal quantum numbers $n$, typically in the
range 30-60. The spatial extent of such highly excited atoms is then
enormous on an atomic scale, exceeding a micron in diameter, and the
transition dipole moments consequently orders of magnitude larger than
those of ground-state atoms or polar molecules. Resonant CP
interactions being proportional to the respective transition dipole
moment squared, a strong enhancement follows. Moreover, an atom
excited to a particular eigenstate with large $n$ is necessarily out
of thermal equilibrium with its environment, and the energy difference
to neighbouring states is typically small compared to $\kB T$ at room
temperature, hence fulfilling the condition for observing oscillating
potential. Remarkably, we recently found that the same property ensures 
that the CP potential in a Rydberg atom is virtually independent of 
temperature from room temperature down to absolute zero \cite{ellingsen10b}.

\paragraph*{Outline of the paper.}
The structure of the paper is as follows. In section
\ref{sec:generalFormalism} we present the general formalism of the
thermal CP interaction and transition rates in a cylindrical cavity,
including the necessary Green tensor. Thereafter, in section
\ref{sec:resonantRadii} we derive the cylinder radii which resonate
with the atomic transition frequency. We begin with the simplest case
of perfectly conducting cavity and then discuss how the optimal radii
deviate from the perfect conductor results when realistic
metallic permittivity functions are employed. We provide simple
formulas for the optimal radii when the permittivity of the cylinder
medium is large. In section \ref{sec:scaling}, we discuss how the
potential and heating rate enhancements scale with the relevant atomic
and cavity parameters, with emphasis on the dependence on the cavity
permittivity. Finally we undertake numerical calculations of resonant
CP potential and heating rates for two example cases: the
$32\mathrm{s}_{1/2}\to 31\mathrm{p}_{3/2}$ transition of Rydberg Rb in
a cylindrical Au cavity at temperature $T=300$K, and a ground state
LiH molecule in a similar cavity.


\section{General formalism}
\label{sec:generalFormalism}

We consider an atomic system (below we study the cases of Rydberg
atoms and polar molecules) with internal energy eigenstates
$|n\rangle$, eigenenergies $\hbar\omega_n$, transition frequencies
$\omega_{mn}=\omega_m-\omega_n$ and dipole matrix elements
$\vect{d}_{mn}=\langle m|\hat{\vect{d}}|n\rangle$, which is prepared
in an incoherent superposition of its energy eigenstates with
probabilities $p_n$. 


\subsection{Thermal Casimir-Polder potential}
\label{SecCP}

The thermal CP force on such a system in an environment of uniform
temperature $T$ was derived in detail in Ref.~\cite{buhmann08}. As
shown, it is conservative in the perturbative limit, 
$\vect{F}(\vect{r}) = -\bm{\nabla}U(\vect{r})$,
where the associated CP potential is given by
\be
 U(\vect{r})=
\sum_{n}p_n U_n(\vect{r}).
\ee
The potential components associated with a given eigenstate $n$ 
splits naturally into a non-resonant contribution $\Unr_n$ and a
resonant one $\Ur_n$,
\be
\label{eq:thermoCP}
  U_n(\vect{r}) = \Unr_n(\vect{r}) + \Ur_n(\vect{r}).
\ee
The non-resonant potential $\Unr_n$ is due to virtual photons and is
reminiscent of that obtained by a dilute-gas expansion of Lifshitz'
formula \cite{lifshitz55}. The resonant contribution $\Ur_n$ is due to
absorption and emission of thermal photons; it is present because the
particle in its ground state is out of thermal equilibrium with its
environment.

The non-resonant potential reads
\begin{equation}
\label{eq:nres}
\Unr_n(\vect{r})=\frac{\kB T}{\ep_0}\jsum
\frac{\xi_j^2}{c^2}\,\tr[{\boldsymbol\alpha}_n (\mi\xi_j) \cdot
 \dyad{G}^{(1)}(\vect{r},\vect{r},\mi\xi_j)]
\end{equation}
where $\mu_0$ is the free-space permeability, $\xi_j = 2\pi j \kB
T/\hbar$ is the $j$th Matsubara frequency, and the prime on the
Matsubara sum indicates that the $j=0$ term is to be taken with half
weight. The atomic/molecular polarizability is given by
\be
  {\boldsymbol\alpha}_n(\omega) 
 = \lim_{\epsilon\to 0}\frac1{\hbar}\sum_k
 \left[\frac{\vect{d}_{kn}\vect{d}_{nk}}{\omega+\omega_{kn}+i\epsilon}
 -\frac{\vect{d}_{nk}\vect{d}_{kn}}
 {\omega-\omega_{kn}+i\epsilon}\right]
\ee
and $\dyad{G}^{(1)}(\vect{r},\vect{r}',\omega)$ is the scattering part
of the classical Green tensor of the geometry the particle is placed
in. Note that dyadic multiplication is implied for products of vectors
without multiplication symbol:
$\vect{A}\vect{B}\equiv\vect{A}\otimes\vect{B}$. For an isotropic
particle, the non-resonant potential components may
be simplified to
\begin{equation}
\label{eq:nresiso}
\Unr_n(\vect{r})=\frac{\kB T}{\ep_0}\jsum
\frac{\xi_j^2}{c^2}\,\alpha_n(\mi\xi_j)
 \tr
 \dyad{G}^{(1)}(\vect{r},\vect{r},\mi\xi_j)
\end{equation}
with
\be
\label{eq:polarizabilityiso}
\alpha_n(\omega)=\lim_{\epsilon\to 0}
 \frac{2}{3\hbar}\sum_k
 \frac{|\vect{d}_{nk}|^2\omega_{kn}}
 {\omega_{kn}^2-\omega^2
-i\epsilon\omega}\,.
\ee

The resonant potential reads
\begin{multline}
\label{eq:res}
\Ur_n(\vect{r})=\mu_0\sum_k\omega_{kn}^2
 \vect{d}_{nk}\cdot\re
 \dyad{G}^{(1)}(\vect{r},\vect{r},|\omega_{kn}|)\cdot\vect{d}_{kn}
 \\
 \times\{\Theta(\omega_{kn})n(\omega_{kn})
 -\Theta(\omega_{nk})[n(\omega_{nk})+1]\},
\end{multline}
where $\mu_0$ is the free-space permeability, and $\Theta(x)$ denotes
the Heaviside step function. The photon number follows the
Bose-Einstein distribution
\be\label{n}
  n(\omega) = 
  \left[\exp\left(\frac{\hbar\omega}{\kB T}\right)-1\right]^{-1}.
\ee
For an isotropic particle, the resonant potential components reduce
to
\begin{multline}
\label{eq:resiso}
\Ur_n(\vect{r})=\frac{\mu_0}{3}\sum_k\omega_{kn}^2
 |\vect{d}_{nk}|^2\tr\re
 \dyad{G}^{(1)}(\vect{r},\vect{r},|\omega_{kn}|)
 \\
 \times\{\Theta(\omega_{kn})n(\omega_{kn})
 -\Theta(\omega_{nk})[n(\omega_{nk})+1]\}.
\end{multline}


\subsection{Environment-assisted transition rates}
\label{SecRate}

An atomic system intially prepared in a given energy eigenstate 
$|n\rangle$ and placed in an environment of uniform temperature $T$
will undergo transitions to different eigenstates due to an absorption
and emission of thermal photons. As shown in
Ref.~\cite{buhmann08b}, the total rate 
\begin{equation}
  \Gamma_n(\vect{r}) = \Gamma_n^{(0)}+ \Gamma_n^{(1)}(\vect{r})
\end{equation}
for transitions out of state $|n\rangle$ consists of a free-space part
$\Gamma_n^{(0)}$ and an environment-induced part $\Gamma_n^{(1)}$. 
In the perturbative limit, these are given as
\begin{subequations}
\begin{align}
\Gamma_n^{(0)}=& 
\sum_k\frac{|\omega_{kn}|^3|\vect{d}_{nk}|^2}
 {3\pi\hbar\ep_0\hbar}
 \{\Theta(\omega_{nk})[n(\omega_{nk})+1]\nonumber\\ 
 &+\Theta(\omega_{kn})n(\omega_{kn})\}\,
\end{align}
and
\begin{align}
\Gamma_n^{(1)}&(\vect{r})
= \frac{2}{\ep_0\hbar}\sum_k\frac{\omega_{kn}^2}{c^2}\,
  \vect{d}_{nk}\cdot
  \im\dyad{G}^{(1)}(\vect{r},\vect{r},|\omega_{kn}|)\cdot\vect{d}_{kn}
 \nonumber\\
&\times\{\Theta(\omega_{nk})[n(\omega_{nk})+1] 
 +\Theta(\omega_{kn})n(\omega_{kn})\}.
\label{eq:rate1} 
\end{align}
\end{subequations}
The environment-induced rate simplifies for an isotropic particle to
\begin{multline}
\Gamma_n^{(1)}(\vect{r})
=\frac{2}{3\ep_0\hbar}\sum_k\frac{\omega_{kn}^2}{c^2}\,
  |\vect{d}_{nk}|^2\tr\im
 \dyad{G}^{(1)}(\vect{r},\vect{r},|\omega_{kn}|)\\
\times\{\Theta(\omega_{nk})[n(\omega_{nk})+1] 
 +\Theta(\omega_{kn})n(\omega_{kn})\} .
\end{multline}


\subsection{The Green tensor in a cylindrical cavity}
\label{SecGreen}

We are going to study atomic systems placed at position
$\vect{r}=(\rho,\varphi,z)$ inside a circularly cylindrical 
free-space cavity of radius $R$ in a bulk non-magnetic medium 
with permittivity $\ep=\ep(\omega)$ as shown in Fig.~\ref{fig:geom}.
The respective scattering Green can be found in Ref.~\cite{li00}
(see also Appendix A4.2 of \cite{scheel08}): 
\begin{widetext}
\begin{align}
  \dyG^{(1)}(\vect{r},\vect{r},\omega) 
   =& \frac{i}{4\pi}\int_{-\infty}^\infty dq\msum\eta^{-2}
    \left[\rMM\Meo(q)\Moe(-q)\right.\pm\rNM\Noe(q)\Meo(-q)   
    \pm\rMN\Moe(q)\Neo(-q)\notag\\
 &\left.+ \rNN\Neo(q)\Noe(-q) \right]
\label{DGF}
\end{align}
where $\eta = \sqrt{k^2-q^2}$ and $k = \omega/c$. The cylindrical 
vector wave functions inside the cavity are \cite{li00, scheel08,
BookTai71}:
\begin{subequations}
\begin{align}
\label{eq:M}
\Meo(q) =& \left[\mp\frac{m}{\rho}\, J_m(\eta \rho)\sincos m\varphi
\hr \right.
 \left.-\eta J'_m(\eta\rho)\cossin n\varphi \hph\right] e^{iqz}\\
\label{eq:N}
\Neo(q) =& \left[\frac{iq\eta}{k}\, J'_m(\eta \rho)\cossin m\varphi
\hr \mp \frac{iqm}{k\rho}\, J_m(\eta\rho)\sincos m\varphi \hph\right.
  +\left. \frac{\eta^2}{k}\,J(\eta \rho)\cossin m\varphi \hz\right]
 e^{iqz}.
\end{align}
\end{subequations}
\end{widetext}
The compact wave vector notation used here implies
$\mathbf{A}_{{}^e_o}\mathbf{B}_{{}^e_o} =
\mathbf{A}_e\mathbf{B}_e+\mathbf{A}_o\mathbf{B}_o$ etc.,
and the upper (lower) sign in \eqref{DGF} corresponds to upper
(lower) index $e, o$ of the vector wave functions. The reflection
coefficients $r$ can be found from a system of linear equations as
described in Ref.~\cite{li00}. For the single-interface cylindrical
cavity in a bulk medium as considered here, the result for the
diagonal coefficients $r_{M}\equiv \rMM$ and $r_{N}\equiv\rNN$
may be written as
\be
\label{eq:refcoeffdef}
  r_{M,N} = - \frac{H^{(1)}_m(x)}{J_m(x)}\tlr_{M,N},
\ee
with
\be\label{tlr}
  \tlr_\sigma = \frac{A+B_\sigma}{A+B_{D}}, \quad \sigma=M,N,
\ee
and 
\begin{subequations}\label{eq:AB}
\begin{align}
  A=& -m^2(kR)^2(qR)^2(\ep-1)^2,\label{eq:A}\\
  B_M=& x_1^2x^2[\ep \th_1^2x^2 -(\th_1\tj 
      +\ep\th_1\th)x_1x +\th \tj x_1^2],\\
  B_N=& x_1^2x^2[\ep \th_1^2x^2 
      -(\ep\th_1\tj+\th_1\th)x_1x+\th\tj x_1^2],\\
  B_D=& x_1^2x^2[\ep \th_1^2x^2 -(\ep+1)\th_1\tj x_1x+\tj^2x_1^2].
\end{align}
\end{subequations}
Here, $x=\eta R$, $x_1=\eta_1R$, 
$\eta_1=\sqrt{\ep(\omega)\omega^2/c^2-q^2}\,$, $\ep=\ep(\omega)$ is
the permittivity of the cylinder and we have defined the shorthand
quantities
\be
\th=\th(x),\quad\th_1=\th(x_1),\quad
 \tj=\tj(x),\quad\tj_1=\tj(x_1)
\ee
where the reduced Bessel functions denoted $\th(x)$ and $\tj(x)$ are
\begin{subequations}\label{hj}
\begin{align}
  \th(x) =& \frac{H^{(1)\prime}_m(x)}{H^{(1)}_m(x)} 
         = \frac{d}{dx}\ln H^{(1)}_m(x);\\
  \tj(x) =& \frac{J'_m(x)}{J_m(x)} 
         = \frac{d}{dx}\ln J_m(x).
\end{align}
\end{subequations}
Note that $\tj$ is a real function for real arguments whereas $\th$ is
complex for real arguments. Explicit knowledge of the off-diagonal
reflection coefficients is not required; for our purposes, it is
sufficient to note that $\rMN=\rNM$ (cf.\ Refs.\ \cite{scheel08,li00}
for details).

Substituting the vector wave functions~(\ref{eq:M}) and (\ref{eq:N}) 
into Eq.~(\ref{DGF}), making use of $\rMN=\rNM$ and the fact that
odd functions of $q$ do not contribute to the integral, 
the Green tensor is found to take the diagonal form 
\begin{widetext}
\begin{align}
\dyG^{(1)}(\mathbf{r},\mathbf{r},\omega)
  =&\frac{i}{2\pi}\int_{0}^\infty 
  dq \msum \left\{\left[\frac{m^2}{\eta^2\rho^2}\,J_m^2(\eta\rho) 
  r_{M}  + \frac{q^2}{k^2}\, J_m^{\prime 2}(\eta\rho) r_{N}\right]
  \hr\hr\right.\notag \\
&+\left[J_m^{\prime 2}(\eta\rho) r_{M} 
 + \frac{m^2q^2}{k^2\eta^2\rho^2}\,
  J_m^2(\eta\rho) r_{N}\right]\hph\hph\left.+ \frac{\eta^2}{k^2}\,
  J_m^2(\eta\rho)r_{N}\hz\hz\right\}.
  \label{Gsandwich}
\end{align}
\end{widetext}
The trace of the Green tensor required for isotropic molecules hence
reads
\begin{multline}\label{G}
\tr\dyG^{(1)}(\mathbf{r},\mathbf{r},\omega)
=\frac{i}{2\pi}\int_0^{\infty} 
dq\msum\biggl\{\biggl(r_M + \frac{q^2}{k^2}\,r_N\biggr)\biggr.\\
\times\left.\left[\frac{m^2}{\eta^2\rho^2}\,J_m^2(\eta\rho) 
 + J_m^{\prime 2}(\eta\rho)\right] +r_N \frac{\eta^2}{k^2}\,
 J_m^2(\eta\rho)\right\}.
\end{multline}

\begin{figure}[th]
  \includegraphics[width=3in]{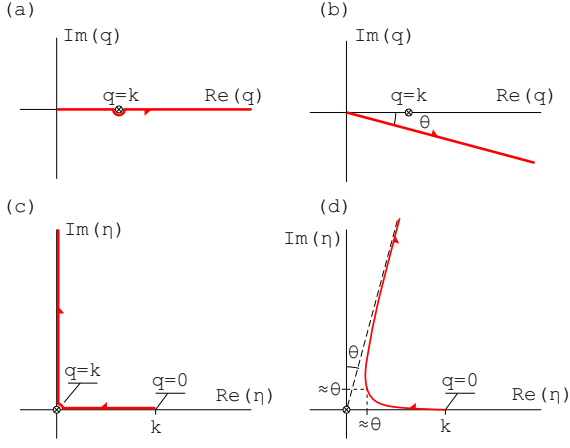}
  \caption{(a) and (b): 
  Rotation of the $q$-integration path by an
  angle $\theta$. (c) and (d): Resulting paths for $\eta$.
}
  \label{fig:path}
\end{figure}

It should be noted that the terms containing $m^2/\eta^2\rho^2$
exhibit third-order poles at $q=k$ where $\eta$ vanishes.
The physically correct treatment of this pole consists of adding a
small imaginary part to the free-space wave vector
$k=\omega/c+i\delta$ and performing the limit $\delta\to 0$. In this
way, the pole is circumvented from below as $q$ is integrated along
the positive real axis, see Fig.~\ref{fig:path}(a). Correspondingly,
$\eta$ is integrated along a part of the real axis, around the
singularity at the origin and along the positive imaginary axis,
Fig.~\ref{fig:path}(c).
Recall that in planar geometries the Green tensor at real frequencies
may be separated into a propagating part ($q<k$) which exhibits an
oscillating behaviour due to interference of incoming and reflected
photons and an evanescent part ($q> k$) which falls off monotonously
away from a surface \cite{ellingsen09a}. In the present case, however,
such a distinction cannot be made due to the pole at the separation
point $q=k$; the individual propagating and evanescent parts would
diverge.  

For numerical purposes, the original integration contour is
unfavourable since it involves subtraction of almost equal numbers, 
associated with a considerable loss of accuracy. Instead, we rotate
the contour of the integral over the variable $q$ to lie along a line
at a small negative angle $\theta$, below the real axis as shown in
Fig.~\ref{fig:path}(b). As shown in Fig.~\ref{fig:path}(d), the
corresponding path for $\eta$ is slightly shifted with respect to the
orginal one but still contained within the first quadrant: When
$\theta\ll 1$, the point of closest approach to the pole at $\eta=0$
is approximately $\theta(1+i)$ and the integration path thence
approaches an asymptote at an angle $\pi/2-\theta$. 

We will show in Sec.~\ref{sec:radii} that all the poles of the
reflection coefficients lie below the real $x$ axis when
$|\ep|<\infty$, approaching the real axis from below in the perfect
conductor limit. There are no other possible singularities in the 
integrands in Eqs.~(\ref{Gsandwich}) and (\ref{G}), so the area
contained between the original and deformed integration contours for
$\eta$ is free of poles. As the intergands vanish along a path at
imaginary infinity connecting the two paths, they are equivalent by
virtue of Cauchy's theorem as long as $\theta < \pi/2$; we choose
$\theta=0.1$ rad. An additional benifit of the rotated integration
path is the fact that the integral becomes exponentially convergent.

The Green tensor becomes particularly simple on the cavity
axis, $\rho=0$ where only a few of the functions $J_m(\eta\rho)$, 
$J_m'(\eta\rho)$ are different from zero. We have for 
$m=0,1,2,...$
\begin{subequations}
\begin{align}
  \frac{m^2}{\eta^2\rho^2}\,J^2_m(\eta\rho) 
  \to& 0, \fourth, 0, 0,...\\
  J^{\prime 2}_m(\eta\rho) \to& 0,\fourth, 0,0,...\\
  J_m^2(\eta\rho) \to & 1,0,0,0,...
\end{align}
\end{subequations}
so that
\begin{multline}
  \dyG^{(1)}(\mathbf{r},\mathbf{r},\omega)|_{\rho=0}
  =\frac{i}{8\pi}\int_{0}^\infty dq
  \left[
  \frac{2\eta^2}{k^2}\, r_{N}^{m=0}\hz\hz \right.\\
  \left.+\biggl(r_{M}^{m=1} + \frac{q^2}{k^2}\,r_{N}^{m=1}\biggr)
  (\hr\hr+\hph\hph)\right].
  \label{Midcavity}
\end{multline}
and
\begin{multline}
  \tr\dyG^{(1)}(\mathbf{r},\mathbf{r},\omega)_{\rho=0}
  =\frac{i}{4\pi}\int_{0}^\infty dq\left(\frac{\eta^2}{k^2}\,
  r_{N}^{m=0}\right.\\
  \left.+r_{M}^{m=1} + \frac{q^2}{k^2}\,r_{N}^{m=1}\right).
\label{MidcavityIsotropic}
\end{multline}

Finally, let us briefly discuss the Green tensor at purely imaginary
frequencies $\omega = i\xi$ as required for the nonresonant
potentials~(\ref{eq:nres}) and (\ref{eq:nresiso}). We have $\eta =
i\sqrt{\xi^2/c^2 + q^2}\equiv i\zeta$, so the arguments of the
cylindrical Bessel and Hankel functions appearing in the reflection
coefficients~(\ref{eq:refcoeffdef})--(\ref{eq:AB}) become purely
imaginary. One has
\begin{subequations}
\begin{align}
  J_m(iy) =& i^m I_m(y), \\
  J'_m(iy) =& i^{m-1} I'_m(y),\\
  H^{(1)}_m(iy) =& {\textstyle\frac{2}{\pi}}\,i^{-(m+1)}K_m(y),\\
  H^{(1)\prime}_m(iy) 
 =&{\textstyle\frac{2}{\pi}}\,i^{-m}K'_m(y),
\end{align}
\end{subequations}
hence the reflection coefficients become ($\sigma=M,N$)
\be
\label{rixi}
  r_\sigma(i\xi) 
  = \frac{2i}{\pi}\,(-1)^m\,\frac{K_m(y)}{I_m(y)}\tlr_\sigma(i\xi).
\ee
The reduced reflection coefficients are found from (\ref{tlr}) with
the substitutions $x\to iy=i\zeta R$, $x_1\to iy_1=i\zeta_1R$ with
$\zeta_1=R\sqrt{\ep(i\xi)\xi^2/c^2 + q^2}$, and $\tj \to \ti/i, \th
\to \tk/i$ with reduced modified Bessel functions analogous to
(\ref{hj}): 
\[
  \ti(y)=\frac{I'_n(y)}{I_n(y)}\,;~~\tk(y)=\frac{K'_n(y)}{K_n(y)}\,.
\]
Using the reflection coefficients~(\ref{rixi}), the Green
tensor~(\ref{Gsandwich}) at purely imaginary frequencies is seen to be
purely real and given by
($\kappa=\xi/c$)
\begin{multline}
  \label{GimFreq}
  \dyG^{(1)}(\mathbf{r},\mathbf{r},i\xi)
  =\frac1{\pi^2}\int_0^\infty dq \msum 
  \frac{K_m(\zeta\rho)}{I_m(\zeta\rho)}\\
  \times\left\{\left[\frac{m^2}{\zeta^2\rho^2}\,I^2_m(\zeta\rho)\tlr_M 
  -\frac{q^2}{\kappa^2}\,
  I^{\prime 2}_m(\zeta\rho)\tlr_N \right]\hr\hr\right.\\
  +\left[I^{\prime 2}_m(\eta\rho)\tlr_M 
  -\frac{m^2q^2}{\kappa^2\zeta^2\rho^2}\,
  I^2_m(\zeta\rho)\tlr_N\right]\hph\hph\\
  -\frac{\zeta^2}{\kappa^2}\,I^2_m(\zeta\rho)\tlr_N \hz\hz\biggr\}.
\end{multline}
Correspondingly, its trace reads
\begin{multline}
  \tr\dyad{G}^{(1)}(\rho,\rho,i\xi) =\frac1{\pi^2}\int_0^\infty dq
  \msum\left\{ \left(\tlr_M 
  - \frac{\zeta^2}{\kappa^2}\tlr_N\right)\right. \\
  \times\left.\left[\frac{m^2}{\zeta^2\rho^2}\,I_m^2(\zeta\rho)
  + I^{\prime 2}_m(\zeta\rho)\right]-\frac{\zeta^2}{\kappa^2}\, \tlr_N 
  I_m^2(\zeta\rho)\right\}\frac{K_m(\zeta\rho)}{I_m(\zeta\rho)}.
\end{multline}
Note that for purely imaginary frequencies, the $q$-inte\-gration in
the Green tensor is unproblematic as $\eta$ remains finite along the
real $q$-axis. 


\section{Resonant radii}\label{sec:resonantRadii}

Combining the general expressions for the CP potential given in
Sec.~\ref{SecCP} with the Green tensor of the cylindrical cavity as
laid out in Sec.~\ref{SecGreen}, we can explicitly calculate the full
potential of any particular atom or molecule in a given state placed
in a cavity of given size and material. Such examples, which require a
numerical analysis, will be given in Sec.~\ref{sec:num}. 

Our main intention is to enhance the resonant potential by means of
the cavity. To that end, it is worth recalling the case of planar
cavity: The resonant CP potential of an atom placed between two plane
surfaces is enhanced for certain interplate separations
\cite{ellingsen09b}. For a given transition frequency there exists a
series of such resonant separations corresponding to integer multiples
of the atomic transition wavelength. In such a case, the transition is
(near-)resonant with a standing-wave mode of the planar cavity.
Mathematically speaking, the enhancement results from a closest
matching of the transition frequency with a pole of the scattering
Green tensor. 

Similarly, for the cylindrical cavity we find that for a given
transition frequency there exists a series of discrete radii such that
the transition is near-resonant with one of the cavity modes. In this
section, we explore analytically the structure of these resonances.


\subsection{Perfect reflector}

Let us begin with the limit $|\ep|\to \infty$ of a perfectly
conducting cavity. For large $|\ep|$, the leading order terms of
the coefficients $A$ and $B_{M,N,D}$ in Eqs.~(\ref{eq:AB}) read
\begin{subequations}
\begin{align}
\label{eq:AApp}
  A \sim& -m^2(kR)^2(qR)^2\ep^2 + \order (\ep)\\
\label{eq:BMApp}
  B_M \sim& -(kR)^3(\eta R)^3\th_1 \th \ep^{5/2}+ \order (\ep^2)\\
\label{eq:BNApp}
  B_N\sim B_D \sim&  - (kR)^3(\eta R)^3\th_1 \tj \ep^{5/2}
 + \order (\ep^2).
\end{align}
\end{subequations}
Hence, in the limit $|\ep|\to \infty$, Eq.~(\ref{tlr}) simplifies to
\be
  \tlr_M \buildrel{|\ep|\to \infty}\over{\longrightarrow}
 \frac{\th}{\tj}\,;~~~ 
 \tlr_N \buildrel{|\ep|\to\infty}\over{\longrightarrow} 1,
\ee
and so the reflection coefficients~(\ref{eq:refcoeffdef}) become 
\be
\label{eq:perfectReflection}
  r_M\buildrel{|\ep|\to \infty}\over{\longrightarrow} 
  -\frac{H_m^{(1)\prime}(x)}{J'_m(x)}\,; 
~~~ r_N\buildrel{|\ep|\to \infty}\over{\longrightarrow}
-\frac{H_m^{(1)}(x)}{J_m(x)}\,.
\ee

The potential~(\ref{eq:thermoCP}) diverges, and is thus truly
resonant, when at least one of these coefficients has a pole at
$q=0$. Since $x=\eta R=kR$ for $h=0$, it is clear that resonances
occur in the perfectly reflecting limit when either $J_m(kR)=0$ or
$J_m^\prime(kR)=0$, that is, when the radius equals one of the radii
given as
\be\label{eq:perf_cond_res}
  R_{mj}^{(\prime)}= \frac{c}{\omega}\,j_{mj}^{(\prime)}
\ee
where $j_{mj}$ and $j_{mj}'$ are the $j$th zero of $J_m(x)$ and
$J_m'(x)$, respectively (only zeros $j_{mj}>0$ are considered).

For each mode $m$ and each polarisation, there is hence a number of
possible radii $R_{mj}$ leading to a resonant enhancement of the CP
potential. The strongest resonance is that corresponding to the
smallest resonant radius, which is $R'_{11}$, corresponding to the
first zero of $J'_1$, $j'_{11}\approx 1.8411838$. If the dominant
transition $|n\rangle\to|k\rangle$ is a downward one, 
i.e.\ $\omega_{kn}<0$, this resonance corresponds to a potential
minimum which can act as a guiding potential. For an upward transition
(such as will be the case e.g.\ for a ground state molecule), the
strongest resonance which corresponds to a potential minimum in the
cylinder center is $R=R_{11}=R'_{01}$, which happens to be a double
resonance since $j_{11}=j'_{01}\approx 3.8317060$. 
We will analyse these examples further in section \ref{sec:num}.


\subsection{Good conductor}\label{sec:radii}

In reality, any metal has a finite conductivity so that
$|\ep|<\infty$. As we will now show, this results in a shifting of the
values of $kR$ which give poles at $q=0$ away from the zeros of the
Bessel functions. As a consequence, the optimal radii for enhancing
the CP potential or transition rates deviate from their
perfect-conductor values as given by Eq.~(\ref{eq:perf_cond_res}). We
derive approximate formulae for the new optimal radii, valid for good
conductors.

In the following we consider a metal described by the Drude model,
\be\label{Drude}
  \ep(\omega) = 1-\frac{\op^2}{\omega(\omega + \mi\gamma)} 
  = 1-\frac{\op^2}{\omega^2 + \gamma^2} 
  + \mi \frac{\gamma}{\omega}\frac{\op^2}{\omega^2 + \gamma^2}
\ee
where $\op$ is the plasma frequency and $\gamma$ is the relaxation
frequency. For good conductors one has $\op\gg \gamma$, and we assume
this is true in the following. Since we are considering molecules, we
will restrict our interest to low frequencies and therefore also
assume $\omega \ll \op$. Under these assumptions the following is true
\be
\label{eq:cond}
  \re \ep <0; ~~ \im \ep > 0; ~~ |\re \ep |,\im\ep \gg 1.
\ee

Let us consider the reflection coefficients $r_{M,N}$
as given by Eqs.~(\ref{eq:refcoeffdef})--(\ref{eq:AB})
once more. We are looking for the complex resonant value of 
$kR$ which corresponds to a zero of the denominator of $r_{M,N}$ when
$q=0$. $A$ vanishes quadratically for small $q$, 
so we can set $A=0$ in the following and consider the coefficients
$B_{M,N,D}$. The sought value of $kR$ is hence the solution of the
equation $B_D|_{q=0}=0$ \footnote{One may note here that the original
pole due to the zero of $J_m(x)$ in the prefactor
(\ref{eq:refcoeffdef}) is cancelled by the presence of $\tj^2$ in the
denomitor whereas $\tj$ only enters to linear order in the numerator.
This is how the pole is moved to the value of $kR$ which solves
Eq.~(\ref{eq:resonanceEq}).}, which may be written as
\be\label{eq:resonanceEq}
  \frac{\th(\sqrt{\ep}kR)}{\tj(kR)} 
  + \frac{\tj(kR)}{\th(\sqrt{\ep}kR)} 
  = \sqrt{\ep} + \frac1{\sqrt{\ep}}.
\ee
The two roots of this second-order equation in $\tj(kR)$ give all the 
resonances since $B_D$ is the denomiator of both reduced reflection
coefficients for $q=0$. The perfect conductor limit $|\ep|\to \infty$
is easily recovered from this equation, in which case the solutions
$\tj(kR)=\infty$ and $\tj(kR)=0$ are just the solutions
(\ref{eq:perf_cond_res}), i.e., $kR =j_{mn}$ and $kR = j'_{mn}$. 

Approximate solutions to (\ref{eq:resonanceEq}) when $|\ep|$ is large
but finite are straightforward to find. We write
\be\label{etaprime}
  kR = j_{mj}^{(\prime)} + \delta^{(\prime)}
\ee
where $\delta^{(\prime)}$ are small complex numbers. The prime
corresponds to solutions close to a zero of $J'_m$. Solving
Eq.~(\ref{eq:resonanceEq}) to leading order in $\ep^{-1}$ then gives
\begin{subequations}
\begin{align}
  \delta\approx& - \frac{i}{\sqrt{\ep}}\,;\label{eq:deltaAppr}\\
  \delta'=\delta'_{mj}\approx& 
\frac{J_m(j'_{mj})}{J''_m(j'_{mj})}
\frac{i}{\sqrt{\ep}}\,;
\end{align}
\end{subequations}
note that $\th(z) \sim i + \order(z^{-1}),~~ |z|\to \infty$ (cf.\
e.g.\ \cite{BookAbramowitz64} \S9.2). The fraction
$J_m(j'_{mj})/J''_m(j'_{mj})$ is a real and negative number of order
unity which tends asymptotically to $-1$ for large arguments.
For a good conductor~(\ref{eq:cond}), $\ep$ is in the second quadrant
of the complex plane, so $\sqrt{\ep}$ is in the first quadrant
for an absorbing medium, and $i/\sqrt{\ep}$ is in the first
complex quadrant. The shifts $\delta$ and $\delta'$ to the poles 
at $q=0$ lie in the third quadrant.

Similarly, one can show that all poles of the integrand of the 
scattering Green tensor of a well conducting surface are displaced
from the Bessel zeros $\eta R= j_{mj}$ and  $\eta R= j'_{mj}$
by small quantities which lie in the third complex quadrant. This 
means that the poles of the reflection coefficients all lie in the
lower half of the complex $\eta$ plane when $|\ep|<\infty$,
justifying the rotation of $q$-integral path mentioned
in Sec.~\ref{SecGreen} and shown in Fig.~\ref{fig:path}.

For real frequencies, the cavity radius can never be chosen such that
it lies exactly on one of the complex-valued resonances. Instead, we
will derive optimal real radii close to the resonances that maximise
the real or imaginary parts of the Green tensor. As these optimal
radii turn out to be different for the real vs imaginary parts, we
have to distinguish between radii which maximise the resonant
potential~(\ref{eq:res}) and those which maximise the heating
rate~(\ref{eq:rate1}).

\subsection{Optimal radii for enhancing the potential}
\label{sec:radpot}

Consider first a resonance associated with a pole of
$r_N$. We represent the pole in the form~(\ref{etaprime}), 
and let $q$ and $\delta$ be small, so that 
\[
  x =\eta R \approx j_{nj} + \delta - \half\,j_{nj} \frac{q^2}{k^2}\,.
\]
Keeping only leading orders in the small quantities $\delta$, $q^2$
and $1/\sqrt{\ep}$, i.e., using
Eqs.~\eqref{eq:AApp}--\eqref{eq:BNApp} with
\begin{align}
\label{eq:BNApp2}
  B_N\sim& -(kR)^3(\eta R)^3\th_1 \tj \ep^{5/2} 
 +(kR)^4(\eta R)^2\tj^2\ep^2\notag\\
&+ \order(\ep^{3/2}),
\end{align}
the reflection coefficient as given by
\eqref{eq:refcoeffdef}-\eqref{eq:AB} becomes
\begin{align}
\label{rNapprox}
  r_N \approx& \frac{i\sqrt{\ep}\hankel(x)}
  {i\sqrt{\ep}J_m(x) -J_m'(x)} \notag\\
   \approx& -\frac{iY_m(j_{mj})}{J'_m(j_{mj})}
 \frac1{q^2j_{mj}/(2k^2)-\zeta}
\end{align}
with 
\be
  \zeta = \delta+ i/\sqrt{\ep}.
\ee
We have used that $\th_1\approx i$ because of its argument being
large with positive imaginary part (see the asymptotic expansions,
section 9.2 of \cite{BookAbramowitz64}) and noted that
$\hankel(x)\approx iY_n(x)$.

Now we note that in the integrand of the integral (\ref{Gsandwich}),
the term which resonates is the last one, which does not have a
prefactor $q^2$. For this term there is no other resonating structures
in the integrand, and we can simply conclude that close to a strong
resonance 
\begin{multline}
\label{prop}
\Ur_n(\vect{r})\propto \re \dyad{G}(\vect{r},\vect{r},\omega)
\propto \im\int_0^\infty dq r_N\\
  \propto \re \int_0^\infty \frac{dq}{q^2-\zeta}\propto
  \re\sqrt{\frac1{-\zeta}} \propto \im \sqrt{\frac1{\zeta}}\,,
\end{multline}
where we have made the substitution $q\to \sqrt{2/j_{mj}}q/k$ at the
beginning of the second line. Explicitly, we have 
\begin{multline}
  \im\sqrt{\frac1{\zeta}}=-\left|\zeta\right|^{-1/2}\sin\left[\half 
  \arctan\frac{\im\zeta}{\re\zeta}\right] \\
  =\frac{-1}{\sqrt{2}\sqrt[4]{\re^2\zeta+\im^2\zeta}}\,
  \sqrt{1-\frac{\re\zeta}{\sqrt{\re^2\zeta + \im^2\zeta}}}
  \label{eq:camelback}.
\end{multline}
where $\re\zeta=\delta-\im\{\ep^{-1/2}\}$ and
$\im\zeta=\re\{\ep^{-1/2}\}$. Differentiating this result with respect
to  $\delta$, we find that the maximum lies at
\begin{equation}
\re\zeta =-\im\zeta/\sqrt{3}\,,
\end{equation} 
so that
\be\label{eq:etasolution}
  \delta \approx \im\{ \ep^{-1/2}\} - \frac1{\sqrt{3}}\re\{
  \ep^{-1/2}\} .
\ee
Note that $\delta$ is independent of $m$ and $j$ to leading order in
the large parameter $\ep$. The resonant radii associated with
resonances of $r_N$ are hence given by
\be\label{resrN}
kR\approx j_{mj} + \im\{ \ep^{-1/2}\} - \frac1{\sqrt{3}}\re\{
\ep^{-1/2}\}.
\ee

For resonances due to $r_M$, we present the poles of $J_m'(x)$ in the
form (\ref{etaprime}). A virtually identical procedure then leads to 
\be\label{eq:etapsolution}
  \delta'=\delta'_{mj} \approx -\frac{J_m(j_{mj}')}{J''_m(j_{mj}')} 
  \delta.
\ee
Consequently, the respective resonant radii read
\be\label{resrM}
  kR\approx j_{mj}  
  -\frac{J_m(j_{mj}')}{J''_m(j_{mj}')}
  \left(\im\{ \ep^{-1/2}\} - \frac1{\sqrt{3}}\re\{
  \ep^{-1/2}\}\right).
\ee

\begin{figure}[tb]
  \includegraphics[width=3.0in]{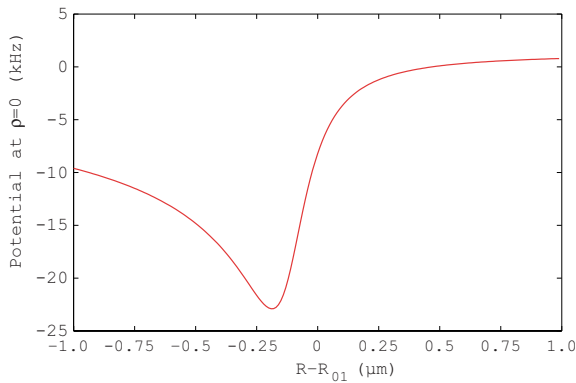}
  \caption{
  The potential of Rydberg Rb in its $32\mathrm{s}$ state at the
  centre ($\rho=0$) of a Au cavity, as a function of radius close to
  the resonance corresponding to the first zero $j_{01}\approx 2.4$ of
  $J_0$. The cavity resonantly enhances the contributino from the
  $32\mathrm{s}_{1/2}\to 31\mathrm{p}_{3/2}$ transition.
  $R_{01}=799.9\upmu$m as given by Eq.\ \eqref{eq:perf_cond_res} is
  the corresponding perfect conductor resonant radius.}
  \label{fig:resonance}
\end{figure}

As a numerical example, take Rb in its $32\mathrm{s}$ Rydberg state
whose dominating 
$|n\rangle = 32\mathrm{s}_{1/2}\to|k\rangle=31\mathrm{p}_{3/2}$
transition has frequency $\omega_{kn}\approx -9.013\times 10^{11}$
rad/s and a cylinder made of Au using $\op\approx 1.4\times 10^{16}$,
$\gamma\approx 5.4\times 10^{13}$ \cite{lambrecht00}. In this case,
the shifts of the maxima away from the Bessel zero for for resonances
due to $r_N$ are found to be $\delta \approx -0.00056$.

In Fig.~\ref{fig:resonance}, we show the potential as a function of
radius close to the smallest resonance at $kR_{01} \approx j_{01}$,
resonating with the smallest downward transition to $31$p. More
details on the specifics of Casimir-Polder potentials on Rydberg atoms
are found in Ref.\ \cite{crosse10} and summarised in
Sec.~\ref{sec:num} below. It is interesting to note that even though
gold is a good conductor whose permittivity is much greater than unity
as assumed above, the shift of the resonant radius away from the
perfect-condutor result is not negligible. We see in Fig.\
\ref{fig:resonance} that the potential at the optimal radius is 
about a factor $2.5$ greater than its value for $kR_{01}=j_{01}$.
The resonant radius is given with excellent approximation by
Eq.~\eqref{resrN} which for this example predicts the maximum at
$R-R_{01}\approx -187$nm.

We note furthermore that the width of the radius resonances is in the
order of $500$nm, which is expected to be well within the accuracy
obtainable for production of pipes with diameters in the order of
hundreds of microns. It is also much wider than surface roughness
amplitudes of good metal surfaces, indicating that the associated
diminishing of the CP-potential enhancement is not expected to be
important. The narrowness of the peaks are thus in the order of one 
three thousandth of the cylinder radius, and we do not expect
observation and utilisation of the resonant behaviour to
be hampered by issues of production accuracy.

\subsection{Optimal radii for enhancing transition rates}

Optimal radii for resonantly enhancing transition
rates in a conducting cavity can be derived in close analogy to the
previous section. We again start with resonances of $r_N$ as
approximated by Eq.~\eqref{rNapprox}. The transition
rates~(\ref{eq:rate1}) close to a resonance are found just as in
\eqref{prop} to be proportional to
\be
\Gamma^{(1)}_n(\vect{r})\propto \im
\dyad{G}(\vect{r},\vect{r},\omega)
\propto \re\sqrt{\frac1{\zeta}}
\ee
where
\begin{multline}
\re\sqrt{\frac1{\zeta}}
=\left|\zeta\right|^{-1/2}\cos\left[\half
 \arctan\frac{\im\zeta}{\re\zeta}\right] \\
=\frac{1}{\sqrt{2}\sqrt[4]
{\re^2\zeta+\im^2\zeta}}\,
\sqrt{1+\frac{\re\zeta}{\sqrt{\re^2\zeta + \im^2\zeta}}}
\label{eq:camelback1}
\end{multline}
The maximum of the above function is again found by differentiation
w.r.t.\ $\delta$. It now lies at $\re\zeta=\im\zeta/\sqrt{3}$, so that
\be\label{eq:etasolutionHR}
\delta
\approx \im\{ \ep^{-1/2}\} + \frac1{\sqrt{3}}\re\{
\ep^{-1/2}\} 
\ee
is again independent of $m$ and $j$ to leading order in $\ep$. The
resonant radii associated with resonances of $r_N$ for enhancing
transition rates
\be\label{resGammarN}
kR \approx j_{mj} + \im\{ \ep^{-1/2}\} + \frac1{\sqrt{3}}\re\{
\ep^{-1/2}\}
\ee
are thus different from the corresponding radii~\eqref{resrN} for
enhancing the CP potential.

As for the potential, the resonances due to $r_M$ are found to be
maximal at
\be\label{eq:etapsolutionHR}
  \delta^{\prime}_{mj} 
 \approx-\frac{J_m(j_{mj}')}{J''_m(j_{mj}')} \delta,
\ee
i.e.,
\be\label{resGammarM}
  kR \approx j_{mj}  
  -\frac{J_m(j_{mj}')}{J''_m(j_{mj}')}
  \left(\im\{ \ep^{-1/2}\} + \frac1{\sqrt{3}}\re\{
  \ep^{-1/2}\}\right).
\ee

\begin{figure}[tb]
  \includegraphics[width=3in]{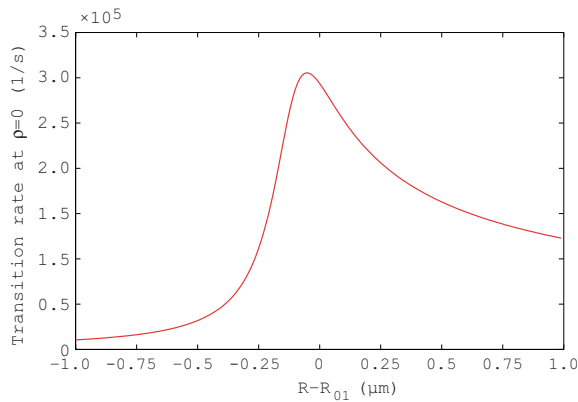}
  \caption{Same set-up as Fig.\ \ref{fig:resonance}, but for the
  cavity-assisted transition rate from state $32\mathrm{s}$ to
  $32\mathrm{p}_{3/2}$. 
}
\label{fig:resonanceHR}
\end{figure}

The resonant enhancement of the total transition rate out of the
$32\mathrm{s}$ state of Rydberg Rb placed at the centre of a Au
cavity is shown in Fig.~\ref{fig:resonanceHR}, again for the $j_{01}$
resonance of $r_N$. The true optinal radius is seen to be well
aproximated by Eq.~(\ref{resGammarN}), it is smaller than the
perfect-conductor value by $0.004$\%. The optimal radii for enhancing
potential vs rate thus differ noteably by about $192~\mathrm{nm}$.
 
\section{General scaling properties}\label{sec:scaling}

In this section we disucss the scaling properties of resonant thermal
CP forces, i.e., their dependences on the relevant molecular,
material, thermal and geometric parameters. These were discussed in
detail for the case of a planar cavity in Ref.~\cite{ellingsen09b};
and many of the results remain valid also in the cylindrical geometry:
The resonant potential corresponding to a dipole transition from state
$n$ to state $k$ is proportional to the absolute square of the
transition dipole moment, $|\vect{d}_{kn}|^2$. Likewise, the resonant
potential scales with temperature just as in the planar case;
proportional to the photon number $n(|\omega_{kn}|)$. For temperatures
larger than the transition frequency the temperature dependence is
thus approximately linear, $\Ur(\rho) \propto T,~ \kB T \gg \hbar
\omega_{kn}$. In the opposite limit the resonant potential is
exponentially suppressed:
$\Ur(\rho) \propto \exp(-\hbar\omega_{kn}/\kB T),~ \kB T \ll \hbar
\omega_{kn}$.

\begin{figure}[tb]
  \includegraphics[width=3in]{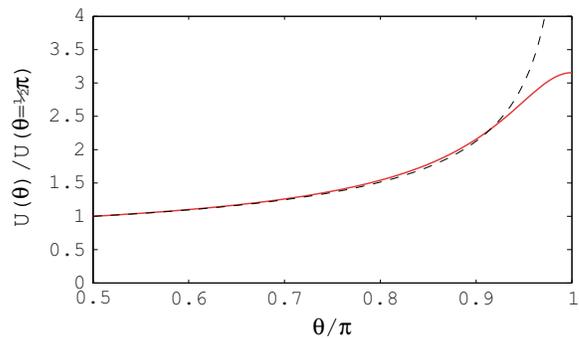}
  \caption{The dependence of the peak potential at resonance on 
  $\theta = \arg(\ep)$. The solid graph is $\Ur$ calculated for a 
  $32$s Rydberg atom whose $32\mathrm{s}\to 31\mathrm{p}$ transition
  is in resonance with the $j_{01}$ zero of $J_0(x)$, varying the
  phase of $\ep$ throughout the physical range but keeping $|\ep|$
  constant and equal to its value using Eq.~(\ref{Drude}). The
  dashed graph is the theoretical curve based on the approximate
  scaling of Eq.\ \eqref{scaling}. Both functions have been
  normalized by their values at $\arg\ep=\pi/2$.}
  \label{fig:theta}
\end{figure}

On the contrary, the scaling of the potential with the permittivity
of the cavity walls depends strongly on the geometry. For the
cylindrical cavity, this scaling follows immediately from the
discussion of the resonant radii in Sec.~\ref{sec:radpot}. As shown by
Eq.~(\ref{prop}), the resonant potential is proportional to
$\sqrt{1/\zeta}$. This function~(\ref{eq:camelback}) has its maximum
at $\re\zeta=-\im\zeta/\sqrt{3}$, where it takes the value
$\sqrt[4]{3}/(2\sqrt{2\im\zeta})$ with $\im\zeta=\re\{\ep^{-1/2}\}$. 
To leading order in the large quantity $\ep$, the potential for
near-resonant radius hence scales as 
\be\label{scaling}
  \Ur_n(\vect{r})\propto \frac{|\ep|^{1/4}}{\sqrt{\cos[\half
\arg(\ep)]}}.
\ee
Note that this scaling is an improvement with respect to the
enhancement achievable in a planar cavity for which we found the
potential to be proportional to $\ln\ep$ \cite{ellingsen09b}. The
resonant enhancement in the above approximation diverges as $\arg
(\ep)\to \pi$. A full numerical investigation shows that the potential
in fact remains finite in this limit, cf.\ Fig.~\ref{fig:theta}. In
the displayed example for the $\arg (\ep)$-dependence of the potential
of Rydberg Rb in its $32\mathrm{s}$ state, the value of the potential
changes by more than a factor of three as the phase of $\ep$ is varied
from $\pi/2$ (purely imaginary permittivity) to $\pi$ (purely real
permittivity). This implies that decreasing the dissipation rate of
the cavity material can increase the enhancement significantly. Note
that with $\gamma\approx 5.4\times 10^{13}$ \cite{lambrecht00}, the
actual phase of Au at the relevant transition frequency is close
to $\pi/2$.

The shown dependence of the resonant potential on the phase of the
permittivity is also interesting in light of the thermal anomaly of
the Casimir effect for metals \cite{brevik06, BookBordag09}. This
dispute centers in an essential way on the description of the
dissipation of the metal: employing the standard Drude model
(\ref{Drude}) with measured optical dissipation data gives a different
prediction of the force at high temperature than that using a
non-dissipative plasma model in which one sets $\gamma = 0$ at the
outset. Experiments appear to favour the latter approach
\cite{BookBordag09, decca06}. The cavity enhancement of the CP
potential as a related quantum vacuum effect is good system for
investigating the thermal anomaly further: As Eq.\ \eqref{scaling}
shows, the potential at resonance using a plasma model predicts a
signal more than three times that calculated for the Drude model
(note that $|\ep|$ is also larger using a plasma model, adding to the
relative difference in prediction). 

Finally, let us consider the scaling of the potential with the
transition frequency. The frequency infuences the
potential~(\ref{eq:res}) in three ways: Firstly, there is a prefactor
$\omega_{kn}^2n(|\omega_{kn}|)$, independently of the particular
cavity geometry considered. Secondly, there is an additional factor of
$\omega_{nk}$ because the magnitude of oscillations fall of as 
$J_m^2, J_m^{\prime 2}\sim 1/(R-\rho)$ away from the cavity walls and
the resonant cavity radius is in turn proportional to $\omega_{kn}$.
This geometric frequency-dependence is closely similar to that found
for a planar cavity \cite{ellingsen09b}; 
it is a consequence of the general scaling law for the Green tensor as
established in Ref.~\cite{RefScaling}. 
Thirdly, the $\ep$-scaling introduces an additional frequency
dependence. For $\omega_{kn}\ll\gamma$, Eq.~(\ref{scaling}) leads to a
$\sqrt[4]{1/\omega_{kn}}$ scaling. Combining these three effects, the
peak resonant potentials scale as
$\omega_{kn}^{2.75}n(|\omega_{kn}|)$. For comparison, the scaling
$\omega_{kn}^3n(|\omega_{kn}|)$ for a planar cavity is slightly
stronger due to the much weaker $\ep$-dependence.

For atoms in highly excited Rydberg states (cf.~Sec.~\ref{sec:num}
below), the CP potential is dominated by transitions to neighbouring
states. For these transitions, the frequencies and dipole matrix
elements depend in a simple way on the principal quantum number $n$
of the initial atomic Rydberg state: For sufficienty large $n$, the
transition frequencies can be given as
$\omega_{kn}\!=\!2\mathrm{Ry}/(\hbar n^3)$ ($\mathrm{Ry}$, Rydberg
energy) while the dipole moments scale as $n^2$. Combining this with
the dependencies discussed above, one finds that the CP potential of a
Rydberg atom scales as 
\begin{align}
  \Ur_n&\propto n^{-4.25}n(\omega_{kn})\notag \\
  &\propto \left\{
  \begin{array}{cl}
    n^{-1.25}, & n\gg n_T;\\
    \exp\left[-(n_T/n)^{3}\right]& n\ll n_T,
  \end{array}\right.
\end{align}
where we have introduced a characteristic thermal principal quantum
number
\be
  n_T = \left(\frac{2\mathrm{Ry}}{\kB T}\right)^\frac{1}{3}.
\ee
The maximum potential is found for states with a principal quantum
number around $n\sim n_T$. At room temperature, $n_T\approx 10.2$,
hence the maximum potential is found for principal quantum numbers
below the Rydberg range. For this reason we have chosen a low Rydberg
state, $32$s, for our numerical examples above and in the following.

\section{Numerical results}\label{sec:num}

We are now going to present numerical studies of the CP potential
inside a cylindrical cavity based on the exact formulas presented in
Sec.~\ref{sec:generalFormalism}. Due to the complexity of the
formulaic apparatus for the case of the cylinder, it is necessary to
first ascertain the correctness of the numerical calculations. As a
numerical benchmark, we 
verified
that for positions sufficiently close
to the cylinder wall ($R-\rho\ll R$) the potential tends
asymptotically to that outside a half-space (c.f.\ e.g.\
\cite{ellingsen09a}). 

\begin{figure*}[tb]
  \includegraphics[width=7in]{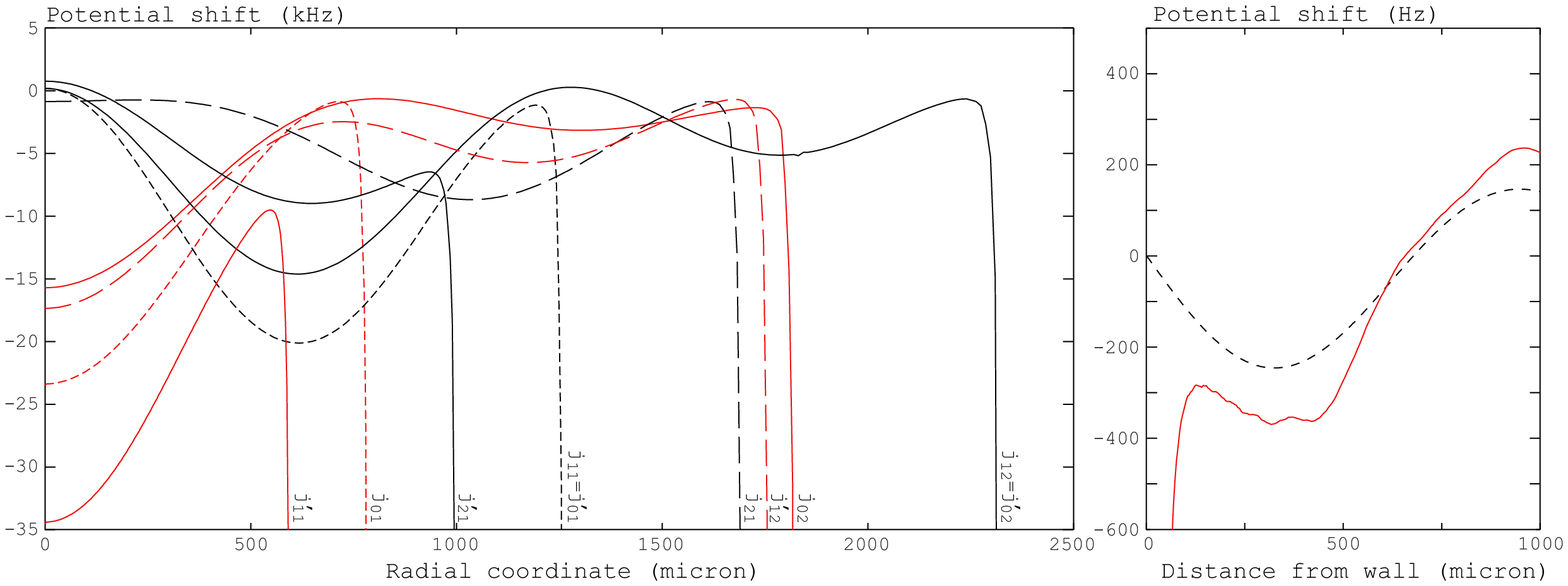}
  \caption{
  (Left panel) Casimir-Polder potential at the eight smallest  
  resonant radii acting on isotropic Rb in the state $32$s, whose
  dominating transition to  $31$p$_{3/2}$ resonantes with various
  modes of a gold cavity described by Eq.\ \eqref{Drude}.  
  The radii are chosen according to Eq.\ \eqref{resrN} ($j_{01}$,
  $j_{02}$, $j_{11}=j'_{01}$, $j_{12}=j'_{02}$, and  $j_{21}$)
  and Eq.\ \eqref{resrM} ($j'_{11}$, $j'_{12}$, $j'_{21}$).
  (Right panel) 
  The unenhanced potential outside a plane half-space. For
  comparison, the oscillating part of the shift due to the transition
  to $31$p$_{3/2}$ alone is shown as a dotted line.
\label{fig:R32graphs}}
\end{figure*}

\begin{figure*}[tb]
  \includegraphics[width=7in]{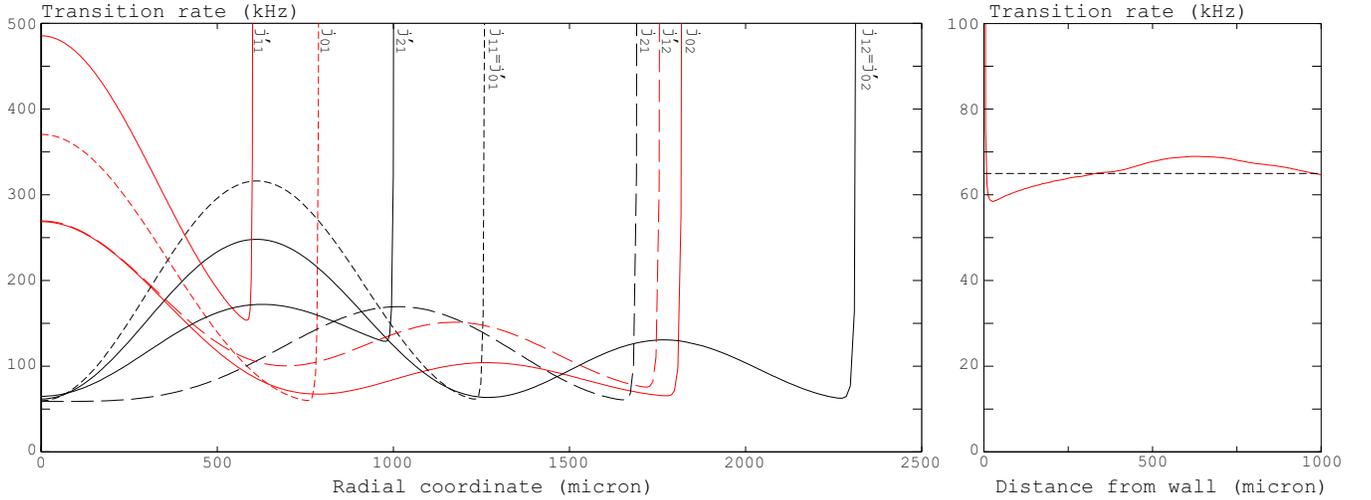}
  \caption{
  (Left panel) Enhanced transition rates at the eight smallest
  resonant radii acting on isotropic Rb in the state $32$s, whose
  dominating transition to $31$p$_{3/2}$   resonates with various
  modes of in a gold cavity described by Eq.\ \eqref{Drude}.  
  The radii are chosen according to Eq.\ \eqref{resGammarN} ($j_{01}$,
  $j_{02}$, $j_{11}=j'_{01}$, $j_{12}=j'_{02}$, and $j_{21}$) and 
  Eq.\ \eqref{resGammarM} ($j'_{11}$, $j'_{12}$, $j'_{21}$).
  (Right panel) 
  The unenhanced shift outside a plane half-space. 
  The free space transition rate is shown as a dotted line.  
}
  \label{fig:HR32graphs}
\end{figure*}

We are interested in an observable resonant enhancement of the
potential predicted for the radii derived in
Sec.~\ref{sec:resonantRadii}. As clear from the previous
Sec.~\ref{sec:scaling}, this requires a cavity made of out a good
conductor and an atomic system whose transition frequencies lie close
to the peak of the thermal spectrum
$\omega_{kn}^{11/4}n(|\omega_{kn}|)$ at room temperature and whose
respective dipole matrix elements are large. Rydberg atoms with their
enormous matrix elements and relatively small transition frequencies
fulfill both of these requirements.

In practice, Rydberg states with principal quantum number $n$ in the
range 30 to 50 can readily be prepared using standard Rydberg lasers.
As shown at the end of Sec.~\ref{sec:scaling},
the optimal choice of $n$ for measuring the resonant enhancement is
smaller than the standard Rydberg regime, around $n=10$: 
While transition dipole moments increase with higher $n$, transition
frequencies of the dominant transition decrease further away from the
optimal frequency value $\sim \kB T$, resulting together in a low
optimum $n$. As a compromise, we use the value $n=32$ for our
numerical examples, being a level in the lower part of the Rydberg
spectrum while still being readily available with standard equipment. 
In all calculation the temperature was $300$K, although in all the 
cases considered herein there is virtually no discernable temperature 
dependence (c.f.\ also \cite{ellingsen10b}).

The resonant potential is calculated only for the dominating
transition to $31$p$_{3/2}$ -- transitions to higher and lower levels
contribute significantly only in the non-retarded regime close to the
cylinder walls which we are not interested in in the present
investigation (but see \cite{crosse10} for details). Results are
shown in Fig.\ \ref{fig:R32graphs} for the first eight resonances
of the coefficient $r_N$ (corresponding to zeros of $J_m(x)$), and 
$r_M$ (corresponding to zeros of $J'_m(x)$). Note how some radii
resonate both with $r_M$ and $r_N$ since $J'_0(x)=-J_1(x)$. 

For the considered case of a resonantly enhanced downward transititon,
we observe that a potential at the resonant radius corresponding 
to $j^{(\prime)}_{mj}$ has $j$ local minima. Potentials for $j_{mj}$
have minima on the cylinder axis for even $m$ and maxima for odd $m$.
For the $j^\prime_{mj}$ resonances, the situation is reversed. The
double resonances are dominated by the $j_{mj}$ contribution, so the
first of the two rules applies. For a given $m$, the maximum potential
depth decreases with $j$.

One might expect that smaller radii give the deeper potential minima,
based on the fact that the amplitude of oscillations decrease away
from a boundary (outside a planar half-space the oscillation
amplitude decreases proportional to inverse distance
\cite{ellingsen09a}). Unlike the planar case \cite{ellingsen09b},
however, this  is not true in general. The lowest Bessel zeros are
$j'_{11}\approx 1.8412$, $j_{01}\approx 2.4048$, $j'_{21}\approx
3.0542$ and $j'_{01}=j_{11}\approx 3.8317$. The two first on this list
indeed correspond to the two largest potential extrema in the same
order, but $j'_{21}$ in fact represents the shallowest of the eight
enhanced potentials considered in Fig.~\ref{fig:R32graphs}. 

\begin{figure}[tb]
  \includegraphics[width=2.8in]{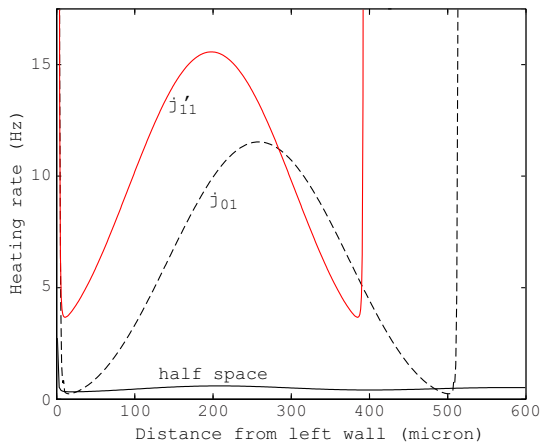}
  \caption{
  Comparison of enhanced and non-enhanced heating rate of ground state
  LiH. The heating rate outside a half-space \cite{buhmann08b} shown
  together with the two highest cylinder resonances (radii from
  \eqref{resGammarN} and \eqref{resGammarM}) along a central cylinder
 cross-section.
}
  \label{fig:LiHgraphs}
\end{figure}

The corresponding enhancement of the transition rate from $32$s to
$31$p$_{3/2}$ is shown in Fig.~\ref{fig:HR32graphs} where the cylinder
radii are picked according to Eq.~\eqref{resGammarN} and
Eq.~\eqref{resGammarM} as appropriate. The figure thus shows the
maximally enhanced transition rate between these two levels. The
transition rate is increased by about a factor $7.4$ for the smallest
cavity corresponding to $j'_{11}$ compared to the transition rate in
free space, which for the $32$s state is about $65$ kHz. In
comparison, the transition rate outside a half-space oscillates about
its free-space value within variations of about $10$ kHz in the far
zone, hence showing that the cavity can enhance the oscillating
contribution to the rate alone by about a factor $50$. The lifetime of
the initial state against spontaneous and stimulated decay is thus
reduced from its free-space value of $50$$\upmu\mathrm{s}$ to
$1.4$$\upmu\mathrm{s}$ inside the cylindrical cavity for the $j'_{11}$
resonance. Recall from Sec.~\ref{sec:resonantRadii} that the
potentials and transition rates peak at different radii. In
Fig.~\ref{fig:HR32graphs}, we have chosen the optimal radii for
enhancing the rates. By contrast, the transition rates at the optimal
radii for enhancing potentials (not displayed) are smaller by about a
factor $2$. 

For comparison, we consider the case of the polar molecule LiH.
Polar molecules also exhibit transitions in the frequency regime ideal
for enhancement, but with the respective dipole matrix elements being 
much smaller than those of Rydberg atoms. LiH has a lowest transition
frequency (rotational) of $\omega_{10}=2.79\times 10^{12}$rad/s and
corresponding transition dipole moment $|\mathbf{d}_{10}|^2=
3.85\times 10^{-58}$C$^2$m$^2$. 
As a result of the smaller dipole moment, the CP potential of polar
molecules is much smaller than that of Rydberg atoms. In spite of a
relative enhancement of about a factor $50$-$100$ in a cylindrical 
cavity, the achievable potential depth is still less than $1$Hz, 
which is insufficient for moelcular guiding purposes. 

While a cavity-enhancement of two orders of magnitude is modest for
the CP potential of a polar molecule, it is of interest when regarding
the ground-state heating rates. As Fig.~\ref{fig:LiHgraphs} shows,
the transition to a particular state (in this case the lowest 
rotational state) can be enhanced by a factor of about $30$, 
considerably reducing the lifetime of the rotational ground-state
against heating from $2.1~$s to $0.064~$s.


\section{Conclusions}

We have presented the detailed theory for a particle (atom or
molecule) in an eigenstate inside a cylindrical cavity carved out of a
homogeneous and non-magnetic material. A particle out of thermal
equilibrium with its environment is subject to spatially oscillating
Casimir-Polder forces near surfaces, and we have focused particularly
on a scheme to enhance the oscillating force components by fine-tuning
the cavity radius to resonate with the particle's internal transition
wavelength. A similar oscillating behaviour is observed for the
transition rates between eigenstates. 

Formulas for calculating appropriate radii for maximal enhancement of
potential and rates have been derived. For a perfectly conducting
cylinder, the resonant radii are exactly given as the zeros of the
Bessel function $J_m$ and its derivatives divided by the wave number
of the resonating transition. The optimal radii are slightly shifted
when the cavity is not perfectly conducting, and simple expressions
for the correction have been derived for good conductors. The
corrected optimal radii for enhancing potential vs transition rates 
are now slightly different. 

We have shown how the cavity enhancement scales with the relevant
paramaters of the set-up. In particular, we have paid attention to 
the dependence of the enhancement on the permittivity
$\ep(\omega_{kn})$ of the cavity material We  have shown that the
potential at resonance scales as $U\propto |\ep(\omega_{kn})|^{1/4}$,
which is a noticeable improvement over the planar geometry, for which
the scaling was found to be logarithmic, 
$U\propto \ln\ep(\omega_{kn})$ \cite{ellingsen09b}. A strong
dependence is also found on the complex phase $\arg \ep(\omega_{kn})$
of the permittivity. This is interesting in light of the controversy
surrounding the temperature correction to the Casimir force between
metals, for which the complex phase of $\ep$ for the metal in question
is of the essence. A precision experiment of the potential enhancement
in a cylindrical cavity could be a critical experiment in this
respect.

The cases of a Rydberg atom and a ground state LiH molecule have been
studied numerically, both of which are of experimental and
technological interest. We have found that the deepest potential
minima for Rydberg atoms can be obtained for quantum numbers in the
lower end of the Rydberg regime. With the smallest cavity enhancement
(corresponding to the first zero of $J'_1(x)$), a guiding potential
depth in excess of $30$ kHz is obtainable, which is within the region
of observability of modern experiments. The enhancement factor
obtained is approximately $50$, an order of magnitude better than what
we obtained for a planar gold cavity \cite{ellingsen09b}. For the
polar molecule LiH, the cavity enhancement was found to be
insufficient to bring the potential into the observable regime.
Instead, a considerable enhancement of ground-state heating rates can
be achieved.

\section*{Acknowledgements}

We have benefited from discussions with Professor J.~Fort\'{a}gh.
This work was supported by the UK Engineering and Physical Sciences
Research Council. Support from the European Science Foundation (ESF)
within the activity `New Trends and Applications of the Casimir
Effect' (\texttt{www.casimir-network.com}) is greatfully acknowledged.

\appendix

\end{document}